\documentstyle[11pt]{article}
\def\lsim{\mathrel{\rlap{\raise 2.5pt \hbox{$<$}}\lower 2.5pt}}
\def\gsim{\mathrel{\rlap{\raise 2.5pt \hbox{$>$}}\lower 2.5pt}}
\begin{document}
\bibliographystyle{plain}
\thispagestyle{empty}
\begin{small}
\begin{flushright}
IISc-CTS-1/98\\
hep-ph/9802338\\
\end{flushright}
\end{small}
\vspace{-3mm}
\begin{center}
{\Large
{\bf The Low Energy Expansion for Pion-Pion Scattering and 
Crossing Symmetry in Dispersion Relations}}

\vskip 2.5cm

B. Ananthanarayan\\
Centre for Theoretical Studies, \\
Indian Institute of Science, \\
Bangalore 560 012, India.\\

\bigskip

\vskip 2cm

\end{center}

\begin{abstract}
We show that a suitable setting for comparison of the low-energy
representation for pion-pion scattering
amplitudes,  with dispersive representation for these amplitudes, 
is provided by certain manifestly crossing symmetric dispersion relations.
We begin with a discussion of fixed-t dispersion relations
and discuss the origin of crossing constraints that arise
in this context when we consider resonance saturation with
certain $l\geq 2$ states.  We demonstrate that the approach
advocated here does not require us to enforce such constraints.
Our results are contrasted with those from fixed-t dispersion relations.    
We finally discuss the numerical import of our results.
\end{abstract}

\noindent PACS Number(s):  11.55.Fv, 11.80.Et, 13.75.Lb, 12.39.Fe

\newpage

\setcounter{equation}{0}
\section{Introduction}
Chiral perturbation theory\cite{gl1} provides a low-energy
representation for pion-pion scattering amplitudes that are manifestly
crossing symmetric and have well-known analyticity properties.
In particular, to $O(p^6)$ in the momentum expansion (with the
pion mass $m_\pi^2$ treated as a quantity of $O(p^2)$) the
pion-pion scattering amplitude may be written down in terms of
three functions of single variables $W_I(s),\,
I=0,1,2$\cite{ssf}.   
One may then compute the low-energy expansion in the energies and pion
masses in terms of these functions which have now been
explicitly computed\cite{kmsf1,bcegs1}.  
Furthermore, the detailed knowledge of the amplitude in field
theory with standard power counting\cite{bcegs1,bcegs2}, 
is expected to lead to
accurate predictions for the I=0 S- wave scattering length $a^0_0$.  
This important physical quantity, whose reported value is
$0.26 \pm 0.05$\cite{nagels} has been measured indirectly only
at the level of $20\%$ accuracy and is expected to be measured
to greater precision\cite{hl1}. 

Pion-pion scattering is a problem that is particular
well suited to dispersion relation analysis\cite{gwsv}, 
rigorously established in axiomatic field theory.
One approach has been to write down fixed-t dispersion relations, with two
subtractions which suffice to guarantee convergence as a consequence
of the Froissart bound, for amplitudes of definite iso-spin.
It has been recognized for a long time that crossing symmetry
in dispersion relations brings in many {\it subtle considerations}.
It is possible to enforce crossing symmetry partially to eliminate
the unknown t-dependent functions in favor of the two S- wave
scattering lengths\cite{smr1}.    We note that it is 
the requirement of two subtractions in the dispersion relations
that makes the accurate determination of the S- wave scattering amplitudes
from dispersion relations problematic --- one needs to combine
theoretical inputs with phenomenology in order to extract precise
predictions for these quantities\cite{bcegs1,bcegs2,hl2}.  

The two subtractions in the dispersion relations we consider,
leave the S- and P- wave absorptive parts completely unconstrained\cite{bgn}
by crossing symmetry.
One may write down the total amplitude as (a) a part that is 
saturated only in terms of these absorptive parts, and (b) a part
that is saturated by the $l\geq 2$ absorptive parts.  
In Ref. \cite{ab1} we explictly showed that (a) may be expressed
in a manner whereby the dispersive representation of the
low-energy polynomials in terms of these absorptive parts may
be explicitly produced.  
Here we discuss case (b) at some length.  

Our analysis begins with considering fixed-t dispersion relations.
We must expand the Roy version of fixed-t dispersion relations
saturated with certain $l\geq 2$ absorptive parts, 
which yield the dispersive representation for the low energy
polynomials in the energies and pion masses.  We immediately encounter
problems that arise from the inadequate implementing
of crossing symmetry by fixed-t disperion relations, which we
discuss at some length.  Results analogous to (a) 
as well as (b) with fixed-t dispersion relations have also been discussed
elsewhere\cite{kmsf1,kmsf2}.  The numerical import of these
shows clearly that any attempt to saturate the
dispersion relations with the absorptive parts of
two of the well-known resonances listed by
the Particle Data Group\cite{pdg}, the $I=0,\, l=2$
state $f_2(1270)$(hereafter referred to as the $f$)
and the $I=1,\, l=3$ state $\rho_3(1690)$(hereafter refered to as the $g$)
violates crossing symmetry --- restoration of crossing
symmetry with these would predict a $\pi \pi$ width of
a nearly 110 MeV for the $g$, which falls
short by a factor of almost 3. 

We therefore advocate a different approach to (b),
which overcomes the problems posed by crossing contraints.
This is achieved by considering certain manifestly
crossing symmetric dispersion relations. {\it We compute
with the $l\geq 2$ absorptive parts given by the $f$ and
$g$,  what is the manifestly crossing symmetric dispersive representation
for the low-energy polynomials in this approximation,  
which constitutes the central result
of this work.}  The technique may be easily extended
to other resonances and also to other models of partial waves.
 
The manifestly crossing symmetric dispersion relations 
are of the type first considered  by Wanders\cite{gw1}
in the context of $\pi^0\pi^0$ scattering and 
subsequently by the authors of Ref.\cite{mrw}
for other totally symmetric amplitudes.
The family of dispersion relations
we consider is parameterized in terms of a variable $x_0$;
we concentrate mainly on the special case of $x_0=0$.  
Note that crossing constraints on the $l\geq 2$ absorptive
parts cannot disappear altogether:  whereas for
a fixed value of $x_0$ the representation does not entail
any crossing constraints, as $x_0$ is varied, relations are
implied on the absorptive parts of the $l\geq 2$ absorptive
parts.  However, one of our significant findings is
that the low-energy polynomials of the type we are interest in,
can be produced only with $x_0=0$. 

We finally note that in
a recent work\cite{gw2}, crossing symmetric dispersion relations
have been employed to fix some of the chiral coupling constants
by comparing the Taylor coefficients of the dispersive and chiral
representations of the pion-pion scattering amplitudes at the
symmetry point $s=t=u=4/3$ in the Mandelstam plane.  

In Section 2 we will review the basics of pion-pion scattering
and remark on the structure of the low-energy expansion.  In Section 3
we will consider the form of fixed-t dispersion relations suitable
for saturation with resonances and discuss the deviations from
crossing symmetry.  In Section 4 we will describe the method of
writing down manifestly crossing symmetric dispersion relations
and evaluate the manifestly crossing symmetric low-energy polynomials.
In Section 5 we discuss the numerical import of our work.  In Section 6
our conclusions are presented.

\setcounter{equation}{0}
\section{Pion-Pion Scattering and The Low Energy Expansion}
Pion-pion scattering :
\begin{equation}
\pi^a(p_1)+\pi^b(p_2) \to \pi^c(p_3) +\pi^d(p_4),
\end{equation}
where $a,b,c,d$ label the iso-spin and the $p_i,\, i=1,2,3,4$,
are all incoming momenta,
may be described in terms of a unique function $A(s,t,u)$,
where $s,\, t,\, u$ stand
for dimensionless Mandelstam variables,
\begin{equation}
s=(p_1+p_2)^2/m_\pi^2, \, t=(p_1+p_3)^2/m_\pi^2, \, {\rm and} \,
u=(p_1+p_4)^2/m_\pi^2,
\end{equation}
with $s+t+u=4$.
We shall introduce units of energy squared where ever
necessary in terms of the square of the pion mass, $m_\pi^2$.
The choice of dimensionless squares of energies is very convenient
in producing the low-energy expansion:  we expand out the integrands
of interest in the dispersive representations out in powers of
$m_\pi^2$.

Since the strong interactions conserve iso-spin, it is customary
to write down amplitudes of definite iso-spin in the s-channel:
\begin{eqnarray}
& \displaystyle T^0(s,t,u)=3 A(s,t,u) + A(t,u,s) + A(u,s,t) & \nonumber \\
& \displaystyle T^1(s,t,u)=        A(t,u,s) - A(u,s,t) & \\
& \displaystyle T^2(s,t,u)=        A(t,u,s) + A(u,s,t) & \nonumber
\end{eqnarray}
The scattering lengths $a^I_l$ arise in the threshold
expansion for the partial wave amplitudes ${\rm Re}f^I_l(\nu)=\nu^l
(a^I_l+b^I_l \nu + ...)$, where the partial wave expansion is
given by $T^I(s,t,u)=32 \pi \sum (2l+1) f^I_l(s) P_l((t-u)/(s-4))$,
$\nu=(s-4)/4$ [the normalization of the amplitudes
here differs with that of \cite{mrw} by $32 \pi$].   
The $b_l^I$ are the effective ranges.

We present our
results for the case of saturating the dispersive
integrals with the absorptive parts due to the $\rho(770)$
(hereafter referred to as the $\rho$), $f$ and
$g$ in the narrow width approximation.  
(We consider the well-known case
of the $\rho$ once more, for two reasons.  Firstly,
since the presence of two subtractions in the dispersion
relations of interest does not constrain the absorptive
part of the $\rho$ from crossing symmetry and our
results will be identical whether we consider
fixed-t dispersion relations or crossing symmetric
ones.  Secondly, the scale of the contribution
to the low-energy coupling constants, or alternatively
the $\beta_i$, from the S- and P- wave absorptive parts
extracted from phenomenology and Roy equation fits to
phase shift and elasticity information, is set by the
$\rho$ as it is the dominant phenomenon in the P- wave
channel in the low to medium energy.) 
The quantum numbers for
these states as well as their physical properties 
are listed in Table 1\cite{pdg}.
The s-channel absorptive parts due to these resonances
in the narrow width approximation:
\begin{equation} \label{f2abs}
A^I(x,t)=32 \pi \sum_l (2l+1)\, {\rm Im} f^I_l(x)P_l(1+{2 t \, m_\pi^2
\over x-4 m_\pi^2})
\end{equation}
and the narrow width approximation reading
\begin{equation}
{\rm Im} f^I_l(x)=\sqrt{x\over x-4 m_\pi^2} \pi \Gamma^{\pi\pi}_m M_m
\delta(x-M_m^2)
\end{equation}
Note that in the narrow-width approximation, the partial waves,
despite being functions of only the variable $x$ do exhibit the
correct normal threshold behaviour.  We will have to keep this
in mind when expanding out the integrands in our dispersive representations.
It must also be noted that the arguments of the Legendre polynomials
do contain $m_\pi^2$ since they could contain information of
the external energies, parametrized by the dimensionless Mandelstam
variables.

The low-energy representation for the pion-pion scattering
amplitude $A(s,t,u)$, computed to two-loop accuracy in chiral perturbation
theory, may be written down in terms of three functions of
single variables, $W_I,\, I=1,2,3$:
\begin{small}   
\begin{equation} \label{lowenergexp}
A(s,t,u)=32\pi\left\{\frac{1}{3} W_0(s)+\frac{3}{2}(s-u)W_1(t)
+\frac{3}{2}(s-t)W_1(u)+\frac{1}{2}W_2(t)+\frac{1}{2}W_2(u)
-\frac{1}{3}W_2(s)\right\}
\end{equation}
\end{small}
These functions are known to have only right hand cuts for
$s>4$ and one then has the dispersion relations for these
functions:
\begin{eqnarray}
& \displaystyle W_0(s)=A_0+B_0 s + C_0 s^2 +D_0 s^3 +{s^4\over \pi}
\int_4^\infty {dy\over y^4 (y-s)} {\rm Im}\, W_0(y) & \\
& \displaystyle W_1(s)=A_1+B_1 s + C_1 s^2  +{s^3\over \pi}
\int_4^\infty {dy\over y^3 (y-s)} {\rm Im}\, W_2(y) & \\
& \displaystyle W_2(s)=A_2+B_2 s + C_2 s^2 +D_2 s^3 +{s^4\over \pi}
\int_4^\infty {dy\over y^4 (y-s)} {\rm Im}\, W_2(y) & 
\end{eqnarray}

The coefficients $A_I, \, B_I, \, C_I,\, D_I, \, I=0,2$, and
$A_1,\, B_1,\, C_1$ are the expansion coefficients of the 
functions $W_I$ and $W_1$ evaluated at $s=0$, and the
amplitude $A(s,t,u)$ may be written most generally as:
\begin{equation}\label{lowenergyexp}
\beta_1 + \beta_2 s + \beta_3 s^2 + \beta_4 (t-u)^2 +
\beta_5 s^3 + \beta_6 s (t-u)^2 + {\rm integral}.
\end{equation}
The $\beta_i$ are known combinations of the these
Taylor coefficients.
It has been shown that the integral in the expression above
has the same structure as that obtained from dispersion relations
with two subtractions with the neglect of all but the absorptive
parts of the S- and P- waves.   In particular,
the $\beta_i$ are completely expressible in terms
of the two S- wave scattering lengths and certain
integrals over the S- and P- wave absorptive
parts\cite{ab1}, in this approximation. Nevertheless, the absorptive parts
of the $l\geq 2$ states do contribute to the $\beta_i$'s at
this order, although they do not contribute to the integral
in eq.(\ref{lowenergyexp})\cite{ssf,kmsf1,kmsf2}.  The objective
of the present work is, of course, to compute these contributions
for certain $l\geq 2$ absorptive parts.
 
\setcounter{equation}{0} 
\section{Fixed-t Dispersion Relations}
A very convenient form of dispersion relations has been
found\cite{smr1} which eliminates the unknown functions $\mu_I, \, \nu_I$
in favour of the S- wave scattering lengths $a^0_0$ and $a^2_0$,
where  This (Roy) form is:
\begin{eqnarray}
T^I(s,t)&=& \sum_{I'}
\mbox{$\frac{1}{4}$}(s\,{\bf 1}^{II'} + t\,
C_{st}^{II'} + u\, C_{su}^{II'})\,T^{I'}_s(4,0) \label{royform} \\
 &+&\!\!\int_{4 \, m_\pi^2}^\infty \!dx\,g_2^{II'}(s,t,x)\,A^{I'}(x,0)
+\int_{4 \, m_\pi^2}^\infty \!dx\,g_3^{II'}(s,t,x)\,A^{I'}(x,t)\,
.\nonumber
\end{eqnarray}
where the crossing matrices $C_{st},\, C_{su}$ and $C_{tu}$
are defined below:
\begin{eqnarray}
& C_{st}=\left[
\matrix{ {1\over 3} & 1 & {5\over 3} \cr {1\over 3} & {1\over 2} & -{5\over 6}
   \cr {1\over 3} & -{1\over 2} & {1\over 6} \cr  } \right], \, \,
C_{su}=\left[
\matrix{ {1\over 3} & -1 & {5\over 3} \cr -{1\over 3} & {1\over 2} & 
  {5\over 6} \cr {1\over 3} & {1\over 2} & {1\over 6} \cr  } \right],
  & \nonumber \\
& C_{tu}=\left[ \matrix{ 1 & 0 & 0 \cr 0 & -1 & 0 \cr 0 & 0 & 1 \cr  } \right],
 &
\end{eqnarray}
and $T^I(4,0)=32 \pi a^I_0,$ and will play no role in
the present discussion.
For our purposes, it is convenient to write the kernels in the form
\begin{small}
\begin{eqnarray} 
g_2(s,t,x)&=&-\frac{t \, m_\pi^2}{\pi\, x\,(x-4 \, m_\pi^2)}\,
(u\, m_\pi^2\, C_{st}  + s\, m_\pi^2 \,C_{st}\, C_{tu} )
\left(\frac{{\bf 1}}{x-t \, m_\pi^2}
+ \frac{C_{su}}{x-s\, m_\pi^2-u \, m_\pi^2}\right)\nonumber \\
g_3(s,t,x)&=&-\frac{s\,u \, m_\pi^4}{\pi\,x(x-4 \, m_\pi^2+t \, m_\pi^2)}
\left(\frac{{\bf 1}}{x-s \, m_\pi^2}
+ \frac{C_{su}}{x-u \, m_\pi^2}\right)\;.
\end{eqnarray}
\end{small}
 
The contribution from the $\rho$ to $A(s,t,u)$ 
is unambiguous\cite{eglpr} and corresponds to
precisely the form of the amplitude presented in Appendix C of
Ref.\cite{gl1}, as well as with the
dispersion relation evaluation presented in Ref.\cite{ab2}:
\begin{small}
\begin{eqnarray}
& \displaystyle  48 \pi \Gamma^{\pi\pi}_\rho
\sqrt{M_\rho^2\over M_\rho^2-4 m_\pi^2} {1\over M_\rho (M_\rho^2-4 m_\pi^2)}
m_\pi^4
\left( {t(s-u)\over M_\rho^2-t\,  m_\pi^2} +{u(s-t)\over M_\rho^2-u \,
 m_\pi^2} \right)
&
\end{eqnarray}
\end{small}
Retaining the kinematic factor that ensures normal
threshold behaviour, we write the amplitude out to the required order below as:
\begin{eqnarray}
& \displaystyle A(s,t,u)_\rho={12\pi \Gamma^{\pi\pi}_\rho
\over M_\rho^7}\sqrt{{M_\rho^2\over M_\rho^2-4 m_\pi^2}} m_\pi^4
\left( -32 M_\rho^2 -192 m_\pi^2 + 32 M_\rho^2 s +208 m_\pi^2 s-
\right. & \nonumber \\
& \displaystyle \left. 6 M_\rho^2 s^2
 -52 m_\pi^2 s^2 +2 M_\rho^2 (t-u)^2 + 12 m_\pi^2 (t-u)^2 +
 3 m_\pi^2 s^3 +m_\pi^2 s (t-u)^2 \right) \label{arho}&
 \end{eqnarray}
In order to generate the dispersive representation for the
low-energy polynomial, we merely expand the functions $g_2,\, g_3$
and the Legendre Polynomials out in powers of $m_\pi^2$.  
A general analysis of the framework considered here has been
discussed earlier\cite{kmsf1,kmsf2}
and will be discussed elsewhere as well\cite{hl2}.
 
The contribution of the $f$  to the iso-spin amplitudes
resulting from the fixed-t (FT) dispersion relations, to this order read:
\begin{small}
\begin{eqnarray}
& \displaystyle T^0(s,t,u)^{FT}_f=\frac{40\pi \Gamma^{\pi\pi}_f}{3 M_f^7}
\sqrt{\frac{M_f^2}{M_f^2-4 \, m_\pi^2}} \, m_\pi^4 \left(
-32 M_f^2-192 \, m_\pi^2 -48 M_f^2 s -560 \, m_\pi^2 s  \right. &
\nonumber \\
& \displaystyle
\left. +14 M_f^2 s^2 +204 \, m_\pi^2 s^2
 +2 M_f^2 (t-u)^2+ 12 \, m_\pi^2 (t-u)^2  
-13 \, m_\pi^2 s^3 + 21 \, m_\pi^2 
s (t-u)^2 \right) & \label{f2t0ft} \\
& \displaystyle T^1(s,t,u)^{FT}_f=\frac{40\pi \Gamma^{\pi\pi}_f}{3 M_f^7}
\sqrt{\frac{M_f^2}{M_f^2-4 \, m_\pi^2}} \, m_\pi^4 \left(96 \, m_\pi^2 s
-16 \, m_\pi^2 (t-u) -4 M_f^2 (t-u) s
\right. & \nonumber \\
& \displaystyle 
\left.  -24 \, m_\pi^2 (t-u) s-48 \, m_\pi^2 s^2 
 + 3 \, m_\pi^2 (t-u) s^2- 6 \, m_\pi^2 (t-u)^2 s+6 \, m_\pi^2 s^3 
+\, m_\pi^2 (t-u)^3 \right) & \\
& \displaystyle T^2(s,t,u)^{FT}_f=\frac{40\pi \Gamma^{\pi\pi}_f}{3 M_f^7}
\sqrt{\frac{M_f^2}{M_f^2-4 \, m_\pi^2}} \, m_\pi^4 \left(
-32 M_f^2-192 \, m_\pi^2+2 M_f^2 (t-u)^2-80 \, m_\pi^2 s \right. & \nonumber \\
& \displaystyle \left. -12 \, m_\pi^2 (t-u)^2   + 3 \, m_\pi^2 
(t-u)^2 s+2 M_f^2 s^2 +
 60 \, m_\pi^2 s^2
-7 \, m_\pi^2 s^3 \right) & \label{f2t2ft}
\end{eqnarray}
\end{small}

The contribution of the $g$ to the iso-spin amplitudes
resulting from the fixed-t dispersion relations, to this order read:
\begin{small}
\begin{eqnarray}
& \displaystyle T^0(s,t,u)^{FT}_g=\frac{56\pi \Gamma^{\pi\pi}_g}{ M_f^7}
\sqrt{\frac{M_g^2}{M_g^2-4 \, m_\pi^2}} \, m_\pi^4 \left(
-32 M_g^2 -192 \, m_\pi^2 + 32 M_g^2 s +368 \, m_\pi^2 s \right. &  \nonumber \\
& \displaystyle \left.  - 6 M_g^2 s^2 -132 \, m_\pi^2 s^2 + 
2 M_g^2 (t-u)^2 + 12 \, m_\pi^2 (t-u)^2 
  + 13 \, m_\pi^2 s^3 -9 \, m_\pi^2 s(t-u)^2 
\right) \label{g3t0ft} & 
\\
& \displaystyle T^1(s,t,u)^{FT}_g=\frac{28\pi \Gamma^{\pi\pi}_g}{ M_f^7}
\sqrt{\frac{M_g^2}{M_g^2-4 \, m_\pi^2}} \, m_\pi^4 \left(- 480 \, m_\pi^2 s
-16 \, m_\pi^2 (t-u)  + 12 M_g^2 (t-u) s +72\cdot \right. & 
\nonumber \\
& \displaystyle \left.  
\, m_\pi^2(t-u) s +
 240 \, m_\pi^2 s^2+ 30 \, m_\pi^2 (t-u)^2 s   
 + 3 \, m_\pi^2 (t-u) s^2 -30 \, m_\pi^2 s^3 + \, m_\pi^2 (t-u)^3\right) & \\
& \displaystyle T^2(s,t,u)^{FT}_g=-\frac{1}{2} T^0(s,t,u)^{FT}_g & \label{g3t2ft}
\end{eqnarray}
\end{small}
Note that the factors $\sqrt{M_m^2/(M_m^2-4 m_\pi^2)}, \, m=f,g$ appear in
the expressions for the amplitudes retaining the memory of the normal
threshold behaviour of the partial waves.

It may be seen from the Roy form of the fixed-t
dispersion relations eq.(\ref{royform}), 
that the amplitudes obtained at $O(p^4)$
are manifestly crossing symmetric and thus there are no crossing
constraints at this order.  However, at $O(p^6)$, we immediately
observe the violation of crossing symmetry by considering, say
$T^1(s,t,u)^{FT}_m$, which for $m=f,g$ contains terms that
are even in powers of $(t-u)$.  Note for instance the term
in $T^1(s,t,u)^{FT}_f$,
$$-240 \pi \Gamma^{\pi\pi}_f/(3 M_f^7) \sqrt{M_f^2/(M_f^2-4 m_\pi^2)}
m_\pi^6 (t-u)^2 s$$
and the term in $T^1(s,t,u)^{FT}_g$,
$$840 \pi \Gamma^{\pi\pi}_g/(M_g^7) \sqrt{M_g^2/(M_g^2-4 m_\pi^2)}.
m_\pi^6 (t-u)^2 s$$
One might then
hope that if indeed the width of the $g$ were to be
\begin{eqnarray}\label{finetune}
& \displaystyle
\Gamma^{\pi\pi}_g={2\over 21} \left({M_g\over M_f}
\right)^6 \sqrt{\frac{M_g^2-4 m_\pi^2}{M_f^2-4 m_\pi^2}} \,
\Gamma^{\pi\pi}_f, &
\end{eqnarray}
then these and possibly other
inconvenient terms would be cancelled.  It turns out that
at $O(p^6)$ there is, in fact, only one crossing constraint\cite{kmsf1,kmsf2} which is
the one given by eq.(\ref{finetune}) as we show below.

We may wish to enforce the direct constraint on
absorptive parts written down by Roy, which is a three component
equation, for $I=0,1,2$:
\begin{eqnarray}
& \displaystyle \int_{4 m_\pi^2}^{\infty} \, dx \left[ \sum_{I'}
\left\{ g_2^{II'}(s,t,x)-(C_{tu} \, g_2(s,u,x))^{II'} \right\} A^{I'}(x,0) 
\right. +
& \nonumber \\
& \displaystyle \left.
\left\{ g_3^{II'}(s,t,x) A^{I'}(x,t)-(C_{tu} \, g_3(s,u,x))^{II'}
A^{I'}(x,u) \right\} \right] = 0 & 
\end{eqnarray}

The left hand side of this equation when saturated with the
absorptive part due to the $f$ in the narrow width approximation
and worked out to $O(p^6)$, yields the result
\begin{eqnarray} 
{640 \, \pi \Gamma_f \over M_f^7} \sqrt{{M_f^2\over M_f^2-4 m_\pi^2}} m_\pi^6 s\, t\, u
\, \, \delta^{I1}, 
\end{eqnarray}
while the result for the $g$ yields
\begin{eqnarray}
-{6720\, \pi
 \Gamma_g \over M_g^7} \sqrt{{M_g^2\over M_g^2-4 m_\pi^2}} m_\pi^6 s\,t\, u
\, \, \delta^{I1}. 
\end{eqnarray}
This reproduces the result eq.(\ref{finetune}) if these
two states were to guarantee crossing symmetry at this order.

We define below the combination $(T^0(s,t,u)^{FT}_m-T^2(s,t,u)^{FT}_m)/3$
and define it to be equal to $A(s,t,u)^{FT}_m$ (despite the fact that
crossing symmetry is not respected) to obtain the expressions below:
\begin{small}
\begin{equation}\label{af2ft}
A(s,t,u)^{FT}_f=
{80 \pi \Gamma^{\pi\pi}_f \over 3 M_f^7} \sqrt{{M_f^2\over M_f^2-4
m_\pi^2}} m_\pi^4
s(-8 M_f^2 - 80 m_\pi^2 + 2 M_f^2 s + 24 m_\pi^2 s - m_\pi^2 s^2
+ 3 m_\pi^2 (t-u)^2),
\end{equation}
\end{small}
and
\begin{eqnarray}
& \displaystyle 
A(s,t,u)^{FT}_g={28 \pi \Gamma^{\pi\pi}_g \over M_g^7}\sqrt{{M_g^2\over
M_g^2-4 m_\pi^2}} m_\pi^4
\left(-32 M_g^2 -192 m_\pi^2 + 32 M_g^2 s +368 m_\pi^2 s -
\right. & \nonumber \\
& \displaystyle
\left. 
6 M_g^2 s^2 -132 m_\pi^2 s^2  + 2 M_g^2 (t-u)^2 + 12 m_\pi^2 (t-u)^2
+ 
13 m_\pi^2 s^3- 9 
m_\pi^2 s(t-u)^2 \right) \label{ag3ft} &
\end{eqnarray}

\setcounter{equation}{0} 
\section{Crossing Symmetric Dispersion Relations}
In Ref.\cite{mrw}, a remarkable dispersive
representation is provided for a system of three
totally symmetric amplitudes, $G_0,\ G_1$
and $G_2$ constructed out of
the three iso-spin amplitudes that we are familiar
with.  In particular, $G_0$
is the $\pi^0\pi^0$ scattering
amplitude.  $G_1$ is constructed by considering the
$I=1$ s-channel amplitude and dividing it by $(t-u)$ and then
considering the cyclic sum.  This amplitude has all
the analytic properties of $G_0$ and thus all results
applicable to it are also applicable to $G_1$.  $G_2$ is
analogously defined after perfoming a further division.
[Note that as a result of these divisions, $G_1$ and $G_2$
in principle verify dispersion relations with fewer
subtractions than $G_0$.]

The dispersive representation is parameterized
by a parameter, $x_0$.   The crossing symmetric dispersion
relations of interest here, arise from first considering
dispersion relations in the so-called $x-y$ plane, where
the Wanders variables $x$ and $y$, are homogeneous variables
of the type $x\sim (st+tu+us)$ and $y\sim stu$.  One then
writes down dispersion relations on straight lines in
the $x-y$ plane given by $y=a(x-x_0)$.  A one-parameter
family of dispersion relations results, when $a$ is further
restricted to be a certain function of $x_0$.  In Ref.\cite{mrw}
$x_0$ is considered in the range,
$0\leq x_0 \leq 50.41$, in order to generate
a closed system of partial wave equations for pion-pion
scattering, whose rigorous validity is significantly larger than
the original system proposed in Ref.\cite{smr1}.  
Note that at fixed $x_0$ crossing
symmetry does not impose any constraints on the
absorptive parts of even the higher waves:  constraints
are obtained by comparing results at differing $x_0$.
One may then
write down the dispersion relations for this case and
extract an expansion in powers of
$m_\pi^2$.   We shall first discuss the case of $x_0=0$
at some length and shall say a few words about $x_0\neq 0$.

\subsection{$x_0=0$}
We shall be interested in the treatment of the following
dispersion relations for the three amplitudes of interest,
for $i=1,2,3$.  Furthermore, in the notation of Ref.\cite{mrw},
we would have to supplement the amplitudes $G_i$ with overlines:
$\overline{G}_i$, but for the present work this is redundant and
shall be omitted.
\begin{eqnarray}
& \displaystyle G_i(s,t,u)=
G_i(4,0,0)+\frac{1}{\pi}\int_{4 \, m_\pi^2}^{\infty} \, d\,x
\, {\rm Disc}\, G_i(x,\tau(x,a)) [x-\tau(x,a)] \cdot
& \nonumber \\
& \displaystyle (2x+\tau(x,a)-4 \, m_\pi^2)
\left[{1\over (x-s \, m_\pi^2)(x-t \, m_\pi^2)(x-u \, m_\pi^2)}-
{1\over x^2 (x-4 \, m_\pi^2)} \right], &
\end{eqnarray}
where
\begin{equation} \label{adef}
a= -{stu  \over 4(st+tu+us)} \, m_\pi^2,
\end{equation}
and
\begin{equation}
\tau(x,a)=\frac{1}{2}\left[-(x-4 \, m_\pi^2)+[(x-4 \, m_\pi^2)^2-\frac{16 
a}{x+4a}
(x(x-4 \, m_\pi^2)]^{1/2}\right].
\end{equation}

Note the relations for the discontinuities of the three amplitudes
in terms of the absorptive parts of the three s-channel iso-spin amplitudes:
\begin{eqnarray}
& \displaystyle {\rm Disc}\, G_0(s,t)=\frac{1}{3}(A^0+2 A^2)(s,t) & \nonumber \\
& \displaystyle {\rm Disc}\, G_1(s,t)=
\frac{1}{6 \, m_\pi^2} \frac{3s -4}{(s-t)(2s+t-4)}(2A^0-5A^2)(s,t) + & 
\nonumber \\
& \displaystyle \frac{1}{\, m_\pi^2} \left[{1\over 2t+s-4} -{2t+s-4
\over 2 (s-t) (2s+t-4)} \right] A^1(s,t) & \\
& \displaystyle {\rm Disc}\, G_2(s,t)=-{1\over 2 \, m_\pi^4}{1 \over
(s-t)(2s+t-4)} (2 A^0 - 5 A^2)(s,t) + & \nonumber \\
& \displaystyle {3\over 2 \, m_\pi^4} {3s-4 \over (2t+s-4)(s-t)(2s+t-4)} 
A^1(s,t) & \nonumber
\end{eqnarray}

Furthermore, our treatment shall concern only the absorptive
parts due to $l\geq 2$ states.  As a result, we have the relations:
\begin{eqnarray}
& \displaystyle G_0(4,0,0)=0 & \nonumber \\
& \displaystyle G_2(4,0,0)={3\over 4 \, 
m_\pi^2} G_1(4,0,0) ={1\over \pi}\int_{4 m_\pi^2}^\infty dx
{2x-4 m_\pi^2\over x(x-4 m_\pi^2)} {\rm Disc} G_2(x,0) &
\end{eqnarray}

The s-channel iso-spin amplitudes are now obtained by
employing the relations:
\begin{eqnarray}
& \displaystyle T^0(s,t,u)={5\over 3} G_0(s,t,u) + & \nonumber \\
& \displaystyle {2\over 9} \left[(3s -4)  G_1(s,t,u)\, m_\pi^2 -
{1\over 3}(3s^2-16+6tu)  G_2(s,t,u) \, m_\pi^4 \right] &\\
& \displaystyle
T^1(s,t,u)={(t-u)\over 9}(3  G_1(s,t,u)\, m_\pi^2+(3s-4) G_2(s,t,u)\, 
m_\pi^4)& \\
& \displaystyle T^2(s,t,u)={2\over 3} G_0(s,t,u) - & \nonumber \\
& \displaystyle {1\over 9} \left[(3s -4) G_1(s,t,u) \, m_\pi^2 -
{1\over 3}(3s^2-16+6tu) G_2(s,t,u) \, m_\pi^4 \right] &
\end{eqnarray}

Following the procedure of expanding out the integrands in powers
of $m_\pi^2$ and inserting the narrow-width approximation for the
absorptive parts of the partial waves, we first evaluate 
the contribution of the $f$  to the iso-spin amplitudes
resulting from the crossing symmetric (CS)
dispersion relations, to this order read:
\begin{small}
\begin{eqnarray}
& \displaystyle T^0(s,t,u)^{CS}_f=\frac{40\pi \Gamma^{\pi\pi}_f}{3 M_f^7}
\sqrt{\frac{M_f^2}{M_f^2-4 \, m_\pi^2}} \, m_\pi^4 \left(
-32 M_f^2-192 \, m_\pi^2 -48 M_f^2 s -496 \, m_\pi^2 s +\right. &
\nonumber \\
& \displaystyle
\left. 14 M_f^2 s^2 
+172 \, m_\pi^2 s^2+2 M_f^2 (t-u)^2+ 12 \, m_\pi^2 (t-u)^2
-9 \, m_\pi^2 s^3+ 17 \, m_\pi^2 s(t-u)^2  \right) & \label{f2t0cs} \\
& \displaystyle T^1(s,t,u)^{CS}_f=\frac{40\pi \Gamma^{\pi\pi}_f}{3 M_f^7}
\sqrt{\frac{M_f^2}{M_f^2-4 \, m_\pi^2}} \, m_\pi^4 (t-u)\left(
-16 \, m_\pi^2 - \right. & \nonumber \\
& \displaystyle 
\left. 4 M_f^2  s -24 \, m_\pi^2  s 
 + 3 \, m_\pi^2  s^2 +\, m_\pi^2 (t-u)^2\right) & \\
& \displaystyle T^2(s,t,u)^{CS}_f=\frac{40\pi \Gamma^{\pi\pi}_f}{3 M_f^7}
\sqrt{\frac{M_f^2}{M_f^2-4 \, m_\pi^2}} \, m_\pi^4 \left(
-32 M_f^2-192 \, m_\pi^2 -112 \, m_\pi^2 s +2 M_f^2 s^2 +
 \right. & \nonumber \\
& \displaystyle \left. 76 \, m_\pi^2 s^2 +2 M_f^2 (t-u)^2+12 \, m_\pi^2 (t-u)^2  
-9 \, m_\pi^2 s^3 + 5 \, m_\pi^2 s(t-u)^2 \right)\label{f2t2cs} &  
\end{eqnarray}
\end{small}
The contribution of the $g$  to the iso-spin amplitudes
resulting from the crossing symmetric dispersion relations, to this order 
read:
\begin{small}
\begin{eqnarray}
& \displaystyle T^0(s,t,u)^{CS}_g=\frac{56\pi \Gamma^{\pi\pi}_g}{ M_f^7}
\sqrt{\frac{M_g^2}{M_g^2-4 \, m_\pi^2}} \, m_\pi^4 \left(
-32 M_g^2 -192 \, m_\pi^2  
+ 32 M_g^2 s  
+ 208 \, m_\pi^2 s \right. &  \nonumber \\
& \displaystyle \left.- 6 M_g^2 s^2 -52 \, m_\pi^2 s^2 + 2 M_g^2 (t-u)^2 
+12 \, m_\pi^2 (t-u)^2 + 3 \, m_\pi^2 s^3 + \, m_\pi^2 s (t-u)^2 
   \right) \label{g3t0cs} & 
\\
& \displaystyle T^1(s,t,u)^{CS}_g=\frac{28\pi \Gamma^{\pi\pi}_g}{ M_f^7}
\sqrt{\frac{M_f^2}{M_f^2-4 \, m_\pi^2}} \, m_\pi^4 (t-u) \left(
-16 \, m_\pi^2  + 12 M_g^2  s  +72
\, m_\pi^2  s +  \right. & \nonumber \\
& \displaystyle \left. 
 \, m_\pi^2 (t-u)^2+ 3 \, m_\pi^2  s^2  \right) & \\
& \displaystyle T^2(s,t,u)^{CS}_g=-\frac{1}{2} T^0(s,t,u)^{CS}_g \label{g3t2cs}& 
\end{eqnarray}
\end{small}

From the $I=0,2$ iso-spin amplitudes, we now write down 
\begin{small}
\begin{equation}\label{af2cs}
A(s,t,u)^{CS}_f=
{160 \pi \Gamma^{\pi\pi}_f \over 3 M_f^7} \sqrt{{M_f^2\over M_f^2-4 m_\pi^2}} 
m_\pi^4
s(-4 M_f^2 - 32 m_\pi^2 + M_f^2 s + 8 m_\pi^2 s + m_\pi^2 
(t-u)^2)
\end{equation}
\end{small}
and
\begin{eqnarray}
& \displaystyle 
A(s,t,u)^{CS}_g={28 \pi \Gamma^{\pi\pi}_g \over M_g^7}\sqrt{{M_g^2\over
M_g^2-4 m_\pi^2}} m_\pi^4
\left(-32 M_g^2 -192 m_\pi^2 + 32 M_g^2 s +208 m_\pi^2 s -
\right. & \nonumber \\
& \displaystyle
\left. 
6 M_g^2 s^2 -52 m_\pi^2 s^2 + 2 M_g^2 (t-u)^2 + 12 m_\pi^2 (t-u)^2+ 
3 m_\pi^2 s^3 
+m_\pi^2 s(t-u)^2 \right) \label{ag3cs} &
\end{eqnarray}
The expressions in eq.(\ref{af2cs}) and eq.(\ref{ag3cs}) constitute
some of the most important results in this work.  Also
note the  striking feature that the amplitude
$A(s,t,u)^{CS}_g$ is identical in structure to eq.(\ref{arho})
$A(s,t,u)_\rho$, upto their respective angular momentum
multiplicites of 7 and 3 respectively.

A check to our results arises from considering the sum of
the contributions from the resonances $f$ and $g'$, where $g'$
is a hypothetical resonance whose mass is equal to that of
the $g$, but whose width is tuned to agree with eq.(\ref{finetune}).
The sum of these amplitudes from fixed-t dispersion relations,
{\it viz.}, eq.(\ref{f2t0ft})-eq.(\ref{f2t2ft}) and
eq.(\ref{g3t0ft})-eq.(\ref{g3t2ft}) the latter evaluated
with the width given by eq.(\ref{finetune})
is identical to those from the crossing symmetric dispersion
relations, {\it viz.}, eq.(\ref{f2t0cs})-eq.(\ref{f2t2cs}) and
eq.(\ref{g3t0cs})-eq.(\ref{g3t2cs}) the latter evaluated
with the width given by eq.(\ref{finetune}).  This is due to
crossing constraints being satisfied by the
former in this event.

\subsection{$x_0\neq 0$}
The dispersion relations for $x_0 \neq 0$ for the $G_i$ turn
out to be far more complicated.  Furthermore, the analog of
eq.(\ref{adef}) for non-zero $x_0$ is now:
\begin{equation}\nonumber
a(x_0)=-{stu \over 4(st+tu+us + 16 x_0)} m_\pi^2.
\end{equation}
Furthermore, the are expressions analogous to $\tau(a)$ of
the previous subsection and dispersive representation for
the $G_i$ has been listed explicitly in Ref.\cite{mrw}.  We have
considered these representations and worked out the 
contributions of our resonances at $O(m_\pi^6)$.
Consider for instance, the contribution of the $f$ at $O(m_\pi^6)$
to the low-energy expansion:
\begin{small}
\begin{eqnarray}
& \displaystyle A(s,t,u)^{CS,x_0}_f=
{{160\,\pi \,{\it \Gamma_f}\,{\sqrt{{{{{{\it M_f}}^2}}\over {{{{\it M_f}}^2}-4 m_\pi^2}}}}\,
      {{{\it m_\pi^4}}}\,\left( -4 + s \right) \,s}\over 
    {3\,{{{\it M_f}}^5}}} + & \nonumber \\
    & \displaystyle
    {{160\,\pi \,{\it \Gamma_f}\,
      {\sqrt{{{{{{\it M_f}}^2}}\over {{{{\it M_f}}^2}-4 m_\pi^2}}}}\,{{{\it m_\pi^6}}}\,
      s\,\over 
    {3\,{{{\it M_f}}^7}\,\left( 16 + 8\,s - 3\,{s^2} - 
        {{\left( t - u \right) }^2} + 16\,{\it x_0} \right) }} }\cdot 
      \left( -512 - 128\,s + 160\,{s^2} - 24\,{s^3} + 
        48\,{{\left( t - u \right) }^2} -  \right. & \nonumber \\
        & \left.
        3\,{s^2}\,{{\left( t - u \right) }^2} - {{\left( t - u \right) }^4} - 
        1216\,{\it x_0} + 480\,s\,{\it x_0} - 44\,{s^2}\,{\it x_0} + 
        60\,{{\left( t - u \right) }^2}\,{\it x_0} \right)  &
\end{eqnarray}
\end{small}
While with $x_0=0$, the low-energy polynomial
was produced by the repeated use of the identity $s+t+u=4$,
for the case at hand, the appearance of the parameter $x_0$ prevents the
factorizing of a series in $m_\pi^2$ into a sum of terms
that is a polynomial in $s$ and $(t-u)^2$ for $A(s,t,u)$.
The result above illustrates our remark.  We also 
note that the answer at $O(m_\pi^4)$ is independent of $x_0$
as expected.  Furthermore, it may be easily checked that in
the limit of $x_0\to 0$, we recover our results of the
previous subsection.
It appears, therefore, that only the choice with $x_0=0$
can produce a low-energy polynomial that is manifestly
crossing symmetric.

The analogous result for the $g$ at this order in $m_\pi$
corresponds to the $x_0=0$ answer even at non-zero $x_0$.
The dependence on this parameter manifests itself at the
next order.

\setcounter{equation}{0}
\section{Numerical Results}
We discuss the numerical import of our work in this section.
We first discuss the case of resonance saturation at $O(p^4)$.
It may be seen by inspection that at this order no crossing
constraints arise.  As a result, we may unambiguously compute
the contributions of the $\rho$, $f$ and $g$  respectively.
The primary reason for our consideration of the $\rho$  is that
it sets the scale for the contributions that arise from the
S- and P- wave absorptive parts.  Crossing symmetry dictates
that the presence of the $\rho$ necessarily
implies contributions to the low-energy polynomials of a similar
order of magnitude from the S- waves; here we do not embark on
a detailed numerical analysis.  
We insert the values of the parameters listed in Table 1,
with $\Gamma^{\pi\pi}$ given by $\Gamma_{tot} \cdot BF$, where
$\Gamma_{tot}$ is the total width and $BF$ is the
$\pi\pi$ branching fraction, into the terms of
$O(p^4)$ of eq.(\ref{arho}), eq.(\ref{af2ft}) and eq.(\ref{ag3ft}).
Comparing with eq.(\ref{lowenergyexp}), we isolate the
$\beta_{i}, \, i=1,2,3,4$ and tabulate the results in Table 2.
An inspection of the Table 2 shows
that at $O(p^4)$ the contributions from the $f$ certainly cannot
be neglected even in comparison with the contribution of the 
$\rho$.  The contribution of the $g$ is an order of magnitude
lower to $\beta_i$.  It is then a standard exercise
to translate these results and read off the
contributions of the resonances to the the one-loop chiral
coupling constants, the $\overline{l_i},\, i=1,2$, which
are also tabulated in Table 2.  Pion-pion
scattering also involves the coupling constants $\overline{l_i},\, i=3,4$.
Resonance saturation does not contribute to these coupling constants
at this order.  
It is also instructive to compare our results with those from effective
lagrangian comptuations of Ref.\cite{dt,bgs,hl2} which are
found to be in agreement, when the constraints from the
Froissart bound are enforced on the latter.  

At $O(p^6)$ we compile the contributions of the resonances
of interest to the $\beta_i$.  This comparison is performed between
eq.(\ref{arho}), eq.(\ref{af2ft}), 
eq.(\ref{ag3ft}) and eq.(\ref{lowenergyexp})
and then between eq.(\ref{arho}),
eq.(\ref{af2cs}), eq.(\ref{ag3cs}) and 
eq.(\ref{lowenergyexp}), upon inserting the values of the parameters
listed in Table 1.  This procedure yields the results tabulated
in Tables 3 and 4 respectively.
One sees that the contributions from the fixed-t
and the crossing symmetric dispersion relations have important
and interesting discrepancies.  However, if the width of the
$g$ is tuned to satisfy eq. (\ref{finetune}), 
(we have referred to such a
hypothetical resonance as the $g'$), then the sum of the contributions
of $f$ and $g'$ to each of the $\beta_i$ is the same no matter
if we compute them from the fixed-t or the crossing symmetric 
dispersion relations as indeed it should be.  

It turns out that the requirment that crossing symmetry be restored at
this order by the $f$ and the $g$, eq.(\ref{finetune}), with 
the known masses of the $f$ and $g$  and the width of
the $f$ from Table 1, predicts 
for the $\pi\pi$ width of the $g$ $\Gamma^{\pi\pi}_g$,
the value of 109 MeV.
This is to be compared with its experimental value
of 38 MeV.   It is precisely this problem which led us to consider
the crossing symmetric representations:
the problem of the mismatch between the ``predicted''
and observed widths of the $g$ resonance
is evaded by considering resonance saturation
with manifestly crossing symmetric dispersion relations.  From
Table 4 we can unambiguously read off the contributions of the
$f$ and the $g$ to the low-energy polynomials. {\it Thus the results
of Table 4 constitute the central results in this work.}  
Somewhat improved values for the $\beta_i$ can
be obtained by replacing the narrow-width approximation 
by other phenomenological models of these and higher partial waves.
Since the resonances considered
here essentially dominate the pion-pion phenomenology, the
present work gives a realistic estimate of the contributions of
the relevant partial waves.

\setcounter{equation}{0}
\section{Conclusions}
In this work, we have shown that $l\geq$ states contribute to
the low-energy polynomial in a manner that can be computed explicitly
from the dispersion relations with two subtractions.  Whereas
fixed-t dispersion relations require us to satisfy crossing constraints,
the problem can be avoided through the use of manifestly crossing
symmetric dispersion relations.  We have shown that, since the
width of the $g$ is smaller by a factor of nearly 3 than the requirement
of crossing constraints of the fixed-t dispersion relations, it
is practical to evaluate the contributions of the $f$ and $g$ states
using the crossing symmetric dispersion relations.  
We conclude with the remark that the subtle property of crossing symmetry
in field theory and dispersion relations must be treated with great
care in matters of principle and practice.

\medskip
\noindent {\bf Acknowledgements:}  It is a pleasure to thank 
Prof. H. Leutwyler for discussions.  I also thank P. B\"uttiker and
D. Toublan for a careful reading of the manuscript, and D. Toublan for 
useful correspondence on tensor meson dominance.

\newpage

\noindent{\bf Table Captions}

\bigskip

\noindent 1. Iso-spin, angular momenta, 
masses, widths (in MeV) and $\pi\pi$ branching
fractions used in our numerical results for the $\rho$,
$f$ and $g$ resonances.  

\noindent 2. Contributions of the $\rho$, $f$ and $g$ resonances
to $\beta_{1,2,3,4}$ at $O(p^4)$ and the corresponding
contributions to $\overline{l}_{1,2}$.  We have taken $m_\pi=139.6$ MeV and
$F_\pi=92.4$ MeV.

\noindent 3.  Contributions of the $\rho$, $f$ and $g$ resonances
to $\beta_{1,2,3,4,5,6}$ at $O(p^6)$ from fixed-t dispersion relations.

\noindent 4.  Contributions of the $\rho$, $f$ and $g$ resonances
to $\beta_{1,2,3,4,5,6}$ at $O(p^6)$ from crossing symmetric
dispersion relations.

\bigskip

\newpage

$$
\begin{tabular}{|c|c|c|c|c|c|} \hline 
$ I$ & $ l$ & $ m $ &  $ M $ & $ \Gamma_{tot} $ & $ BF $ \\ \hline
1 & 1 &$\rho$ & 769 & 151 & $100\%$ \\ \hline
0 & 2 & $ f $ & 1275 & 185 & $85\%$ \\ \hline
1 & 3 & $ g $ & 1690 & 160 & $24\%$ \\ \hline
\end{tabular}
$$

\medskip
$$
{\rm Table\, 1}
$$

\bigskip

$$
\begin{tabular}{|c|c|c|c|c|c|c|}\hline
$ m $ & $\beta_1$ & $\beta_2$ & $\beta_3$ & $\beta_4$ & $\overline{l}_1$
& $\overline{l}_2$ \\ \hline
$ \rho $ & $-0.276$ & $0.276$ & $-0.0517$ & $0.0172$ & $-6.27$ & $3.13$ \\ \hline
$ f $ & 0 & $-0.0122$ & $0.00305$ & $0$ & $0.277$ & $0$ \\ \hline
$ g $ & $-0.0030$ & $0.0030$ & $-5.66\, \cdot 10^{-4}$ & $
1.88\, \cdot 10^{-4} $ & -0.069 & 0.034 \\
\hline
\end{tabular}
$$

\medskip
$$
{\rm Table\, 2}
$$

\newpage

$$
\begin{tabular}{|c|c|c|c|c|c|c|}\hline
$ m $ & $\beta_1 $ & $\beta_2 $ & $\beta_3 $ & $\beta_4 $ & 
$\beta_5 $ & $\beta_6 $  \\ \hline
$\rho$ & $-0.330$ & $0.335$ & $-0.0665$ & $0.0210$ & $8.54 \, \cdot 10^{-4}$ &
$2.84 \, \cdot 10^{-4}$ \\ \hline
$f$ & $0$ & $-0.0136$ & $0.00348$ & $0$
 & $ -1.82 \, \cdot 10^{-5} $ & $ 5.47 \, \cdot 10^{-5} $
\\ \hline
$g$ & $-0.00314$ & $0.00326$ & $-6.51 \, \cdot 10^{-4}$ & $1.96\, \cdot 10^{-4}$ &
$8.36\, \cdot 10^{-6}$ & $-5.79 \, \cdot 10^{-6}$ \\ \hline
\end{tabular}
$$

\medskip

$$
{\rm Table\, 3}
$$

\medskip

$$
\begin{tabular}{|c|c|c|c|c|c|c|}\hline
$ m $ & $\beta_1 $ & $\beta_2 $ & $\beta_3 $ & $\beta_4 $ & 
$\beta_5 $ & $\beta_6 $  \\ \hline
$\rho$ & -0.330 & 0.335 & -0.0665 & 0.0210 & $8.54 \, \cdot 10^{-4}$ &
$2.84 \, \cdot 10^{-4}$ \\ \hline
$f$ & $0$ & $-0.0133$ & $0.00333$ & $0$ & $ $0$ $ & $ 3.65 \, \cdot 10^{-5} $
\\ \hline
$g$ & $-0.00314$ & $0.00315$ & $-5.99 \, \cdot 10^{-4}$ & $1.96\, \cdot 10^{-4}$ &
$1.93\, \cdot 10^{-6}$ & $6.43 \, \cdot 10^{-7}$ \\ \hline
\end{tabular}
$$

\medskip

$$
{\rm Table\,  4}
$$
\end{document}